\documentclass[aps,prd,superscriptaddress,nofootinbib]{revtex4}%[]{article}

\pdfoutput=1
\usepackage{amsmath,amssymb,latexsym}
\usepackage{mathrsfs,graphicx,subfigure}
\usepackage{calrsfs}
\DeclareMathAlphabet{\pazocal}{OMS}{zplm}{m}{n}

\graphicspath{
	{images/}
	{.}
}

\usepackage{xspace}	% \xspace command decides whether to add space or not at the end of a text defined by a \newcommand

\usepackage{array} % to align in tables
\usepackage[pdftex,bookmarks,colorlinks]{hyperref}
\hypersetup{colorlinks,%
	citecolor=blue,%
	filecolor=black,%
	linkcolor=green,%
	urlcolor=blue,%
}

\usepackage{cleveref} %for reference to multiple eqs in a single command \cref{1,2,..}
\newcommand{\crefrangeconjunction}{--} %for cref to separate eqs with a dash
\crefformat{equation}{(#2#1#3)}% formatting creef with single arg
\crefrangeformat{equation}{(#3#1#4)\crefrangeconjunction(#5#2#6)}% formatting cref with multiple arg
\crefmultiformat{equation}{(#2#1#3)}%
{ and~(#2#1#3)}{, (#2#1#3)}{ and~(#2#1#3)}% formatting cref with multiple arg with 'and'

\newcommand{\diff}{\operatorname{d}\!}

\newcommand{\trf}{\mathrm{tr}(f)}
\newcommand{\detf}{\mathrm{det}(f)}
\newcommand{\dd}{\mathrm{d}}
\newcommand{\be}{\begin{equation}}
\newcommand{\ee}{\end{equation}}
\newcommand{\ex}[1]{\mbox{e}^{\,\textstyle#1}}

\makeatletter
\def\section{\@startsection {section}{1}{\z@}{-3.5ex plus -1ex minus
		-.2ex}{2.3ex plus .2ex}{\large\sc}}
\def\subsection{\@startsection{subsection}{2}{\z@}{-3.25ex plus -1ex minus
		-.2ex}{1.5ex plus .2ex}{\normalsize\sc}}
\makeatother

\makeatletter
\@addtoreset{equation}{section}

\makeatother

\begin{document}

\title{Geodesic congruences, impulsive gravitational waves\\
and gravitational memory}

\author{Martin O'Loughlin}
\email{martino@fluiditj.com}
\author{Hovhannes Demirchian}
\email{demhov@bao.sci.am}
\affiliation{Ambartsumian Byurakan Astrophysical Observatory, Byurakan, 0213,  Armenia}
	
\begin{abstract}
  In this paper we introduce a new approach to the study of the effects that an impulsive wave, containing a mixture of material sources and gravitational waves, has on a geodesic congruence that traverses it. We find that the effect of the wave on the congruence is a discontinuity in the $\pazocal{B}$-tensor of the congruence. Our results thus provide a detector independent and covariant characterization of gravitational memory. We note some similarities between our results and the study of soft gravitons and gravitational memory on $\pazocal{I}$.
\end{abstract}

\maketitle

%================================================================================

\section{Introduction}

The study of impulsive gravitational waves in the form of null shells was initiated by Penrose and others\cite{Synge:1957zz,Israel66,Penrose:1972xrn}. This topic has recently received renewed attention due to their possible role in the transfer of information from black hole horizons to null infinity. As the black hole horizon is a killing horizon, there is an infinite variety of ways to attach (solder) the black hole interior to the black hole exterior creating a null shell on the horizon \cite{Barrabes:1991ng,Blau:2015nee}. A subclass of these can be shown to correspond to BMS like supertranslations. Furthermore the long studied BMS supertranslations at null infinity of asymptotically flat spaces are linked to the physics of soft gravitons \cite{Weinberg:1965nx,Strominger:2013jfa,He:2014laa} which appear to play an important role in restoring information not seen in the hard gravitons of Hawking radiation \cite{HPS1,HPS2}. In turn the soft gravitons are related to the gravitational memory effect \cite{Strominger:2014pwa}.

Gravitational memory \cite{grav_memory,Grishchuk:1989qa,Bondi:1989vm,Strominger:2017zoo,Ashtekar:2018lor,Donnay:2018ckb,Compere:2018ylh}  is the classical change in nearby geodesics in an asymptotically flat region of space-time as they pass through an outgoing gravitational wave. The study of the effect of a null shell on a time-like congruence that crosses it has been addressed by Barrabes and Hogan \cite{Barrabes_Hogan-book, Barrabes:2001vy}. They calculated the change in the tangent vector and the geodesic deviation vector together with the expansion, shear and rotation upon crossing an impulsive gravitational wave and found a jump in the acceleration of the geodesic and derivatives of the geodesic deviation vector proportional to the stress-energy content and gravitational wave components of the shell. 

To further understand the relationship between gravitons and gravitational memory it is thus important to study the effect of waves on null geodesic congruences, not only as the congruence crosses the wave but also the future evolution of the congruence. In this paper we describe a new exact approach for studying the effect of null shells on null geodesic congruences. This method allows one to easily calculate the change in the $\pazocal{B}$-tensor, which encodes the expansion, shear and rotation of the congruence, upon crossing the shell and its evolution to the future of the shell. We find that the effect of the shell on the congruence, as already observed in the time-like case in \cite{Barrabes:2001vy,Barrabes_Hogan-book}, is a discontinuity in the $\pazocal{B}$-tensor and we will refer to this memory of the passing wave carried by the $\pazocal{B}$-tensor as the $\pazocal{B}$-memory effect, not to be confused with the B-mode gravitational memory. We show how this $\pazocal{B}$-memory is determined by the stress energy and gravitational wave components of the shell. We consider the simplest case of a null shell representing an outgoing gravitational wave and parametrised by a general soldering transformation (a subclass of which are the BMS supertranslations) in Minkowski space, but our method is applicable to any geodesic congruence that crosses a null shell localised on a killing horizon. It is intriguing to note that our formulation of $\pazocal{B}$-memory has much in common with gravitational memory as formulated in \cite{Ashtekar:2018lor}.

In section \ref{sec:memory} we introduce the concept of a $\pazocal{B}$-memory effect as a covariant formulation of gravitational memory. In section \ref{sec:setup} we describe the setup of the problem and give a general description of the suggested approach. A detailed discussion of the approach is carried out in section \ref{sec:application} while in section \ref{sec:behaviour} the detailed behavior of a lightlike congruence is studied. In section \ref{sec:discussion} we discuss our results and their relation to other formulations of gravitational memory, in particular to that reviewed in \cite{Ashtekar:2018lor}.

%================================================================================

\section{$\pazocal{B}$-memory as gravitational memory.}
\label{sec:memory}

The gravitational memory effect is the change in relative velocity between neighboring geodesics after the passing of a gravitational wave - the idea being that the passing of a gravitational wave leaves some ``memory'' in the relative movement of inertial observers. Here we propose a more covariant characterization of this memory effect by considering the effect of an outgoing wave in the form of a null shell on a null geodesic congruence.

To see explicitly how this works we begin with the general construction and notation of \cite{Blau:2015nee}. The impulsive wave (null shell) is confined to a singular null hypersurface $\pazocal N$ which divides the space-time into two domains $\pazocal M^-$ $\bigcup$ $\pazocal M^+$ - the past and future domains - each with its own coordinate system $x^\mu_\pm$. Each domain has its own metric, $g_{\mu\nu}^-$ or $g_{\mu\nu}^+$, together with junction conditions for soldering that relate the two metrics where they meet on the hypersurface $\pazocal N$.
The soldering determines the constituents of the impulsive wave and in the case that $\pazocal N$ coincides with a killing horizon an infinite variety of solderings are allowed \cite{Blau:2015nee} producing an infinite variety of impulsive signals. For explicit calculations we will use the freedom to perform independent coordinate transformations on $\pazocal M^-$ and $\pazocal M^+$ to choose a global coordinate system $x^\mu$ that is continuous across $\pazocal N$ and such that the metric is also continuous
\be
[g_{\mu\nu}] = g_{\mu\nu}^+ - g_{\mu\nu}^- = 0.
\ee
In these global coordinates the hypersurface $\pazocal N$ is defined by the equation $\Phi(x)=0$ with $\Phi(x)>0$ covering the future domain and $\Phi(x)<0$ covering the past domain. 

We will consider a congruence with tangent vector field $T$ transverse to $\pazocal N$ together with the null generator $n$ of the shell, where $T\cdot n=-1$, and to calculate the independent components of the $\pazocal{B}$-tensor $\pazocal{B}_{\alpha\beta} = \nabla_\beta T_\alpha$ \cite{Poisson} we will project it onto the spatial submanifold of the shell defined by a pair of space-like orthonormal vectors $e_A^\alpha$, $A\in(x,y)$ such that $e_A\cdot n=e_A\cdot T = 0$. Furthermore we can and will choose $e_A^\alpha$ to be parallel transported along the congruence, a choice that simplifies the following equations by eliminating the connection from the evolution equation for $\pazocal{B}$. The projection of $\pazocal{B}_{\alpha\beta}$ onto the congruence is
\be
\pazocal{B}_{AB} = e_A^\alpha e_B^\beta \pazocal{B}_{\alpha\beta} = \frac{1}{2}\theta\delta_{AB}+ \sigma_{AB}+\omega_{AB},
\ee
where the expansion, shear and rotation are explicitly given by
\be
\label{eq:BAB}
	\theta = \pazocal{B}^A_A\qquad
	\sigma_{AB} = \pazocal{B}_{(AB)} - \frac12 \theta\delta_{AB}
        \qquad
        \omega_{AB}=\pazocal{B}_{[AB]}.
\ee

The evolution equation for $\pazocal{B}_{AB}$ (with respect to the affine parameter $\lambda$ of the congruence) is
\be
\label{eq:Bevolution}
\frac{\dd \pazocal{B}_{AB}}{\dd \lambda} = -\pazocal{B}_{AC}\pazocal{B}_B^C -R_{AB}
\ee
and
\be
R_{AB} = R_{\alpha\mu\beta\nu}e_A^\alpha T^\mu e_B^\beta T^\nu = \frac{1}{2}\mathcal{R}\delta_{AB} + C_{AB}
\ee
where
\be
\mathcal{R} = R_{\alpha\beta}T^\alpha T^\beta \qquad C_{AB} = C_{\alpha\mu\beta\nu}e_A^\alpha T^\mu e_B^\beta T^\nu
\ee
and $C_{AB}$ is traceless.

In the presence of a null shell the Riemann and Weyl tensors have a term that
is localised on the shell and proportional to a delta function \cite{Barrabes:1991ng}. Thus we separate $\mathcal{R}$ and $C_{AB}$ into their bulk and shell components
\be
\mathcal{R}=\hat{\mathcal{R}} + \bar{\mathcal{R}}\delta(\Phi)\qquad C_{AB} = \hat{C}_{AB} + \bar{C}_{AB}\delta(\Phi),
\ee

In the evolution equation for $\pazocal{B}_{AB}$ the delta function in $R_{AB}$ on the right hand side can only be balanced by a delta function in the derivative of $\pazocal{B}_{AB}$
meaning that the $\pazocal{B}$-tensor must be discontinuous across the shell. This discontinuity is related to the stress-energy and gravitational wave components of the shell as we will see in detail in the following sections. 

The evolution of the rotation is simply given by
\be
\frac{\dd\omega_{AB}}{\dd \lambda} = -\theta \omega_{AB},
\ee
which can be integrated to give
\be
\omega_{AB} = K \ex{-\int_{\lambda_0}^\lambda \theta\dd \lambda^\prime}\epsilon_{AB}.
\ee
We can deduce from this equation that the rotation must be continuous but not necessarily differentiable across the shell as the expansion is at most discontinuous. In particular, and as we will see in detail in the following sections, a zero rotation before the shell and at worst a finite jump in the expansion will result in zero rotation after the shell. This means that a congruence that is hypersurface orthogonal to the past of the shell must also be hypersurface orthogonal to the future of the shell.

Our calculations thus indicate that an alternative and generally covariant formulation of the gravitational memory effect is that there is a discontinuity in the $\pazocal{B}$-tensor of a congruence upon crossing a null shell. In the following sections we will show how to explicitly calculate the evolution of the $\pazocal{B}$-tensor for a congruence that crosses a null shell. 

\section{The setup and proposal}
\label{sec:setup}
Our general construction is applicable to any null shell located on a killing horizon. For simplicity (and without loss of conceptual insight) we will consider in the following sections exclusively the case of a planar null hypersurface (which is obviously a killing horizon) in Minkowski space.

\begin{figure}[h!]
  \center{\includegraphics[width=7cm]{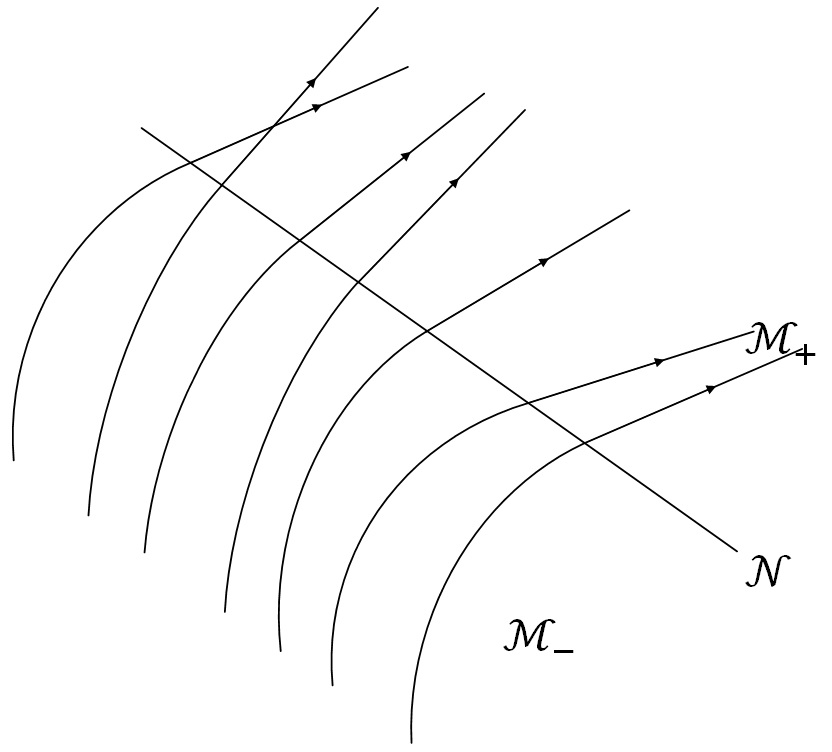}}
  \caption{In continuous coordinates the geodesic vector field is continuous across $\pazocal{N}$. Here we see that the transformed vector field to the past of $\pazocal{N}$ provides the initial conditions for the field to the future and thus the full solution to the geodesic equation.}
  \label{cts-lines}
\end{figure}

To study the evolution of a null geodesic congruence upon crossing a null shell we start directly from the geodesic equation. In continuous coordinates by definition the metric is continuous across $\pazocal N$ while the Christoffel symbols are discontinuous, and the Riemann tensor has a delta function singularity localised on the shell, these properties being directly related to the stress-energy tensor of the shell and explained in detail in \cite{Blau:2015nee}. For the purposes of our calculations we will obtain continuous coordinates across the shell by performing a coordinate transformation on $\pazocal M^-$ while leaving $\pazocal M^+$ in flat coordinates. 

\begin{figure}[h!]
  \center{\includegraphics[width=7cm]{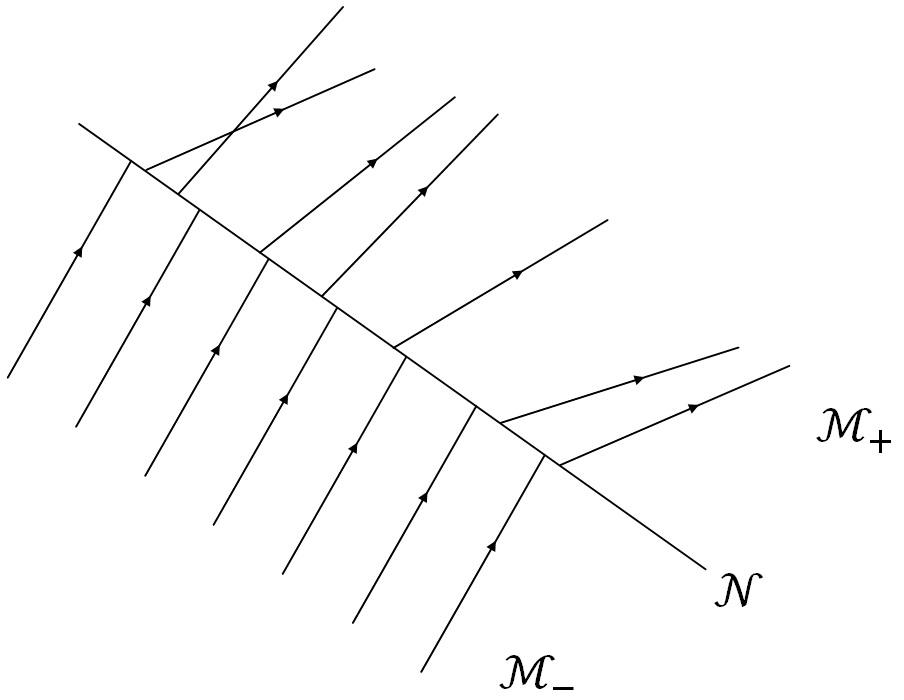}}
  \caption{With flat coordinates to the past and future the soldering transformation leads to a discontinuity across ${\pazocal N}$ in both coordinates and in the geodesic congruence.}
  \label{discts-lines}
\end{figure}

The geodesic equation in the vicinity of the shell is
\be
\ddot{X}^\mu + (\Theta(-\Phi)\Gamma^{-\mu}_{\nu\lambda}+\Theta(\Phi)\Gamma^{+\mu}_{\nu\lambda})\dot{X}^\nu\dot{X}^\lambda =0
\ee
where $\Theta$ is the Heaviside step function. It is clear that non-trivial solutions to this equation may have a discontinuity in the acceleration, but not in the tangent vector $T=\dot{X}$, and thus the geodesic flow lines are $C^1$ across the shell as shown in figure \ref{cts-lines}. Mathematically speaking this means that the geodesic vector on the shell ($T_0$) is uniquely defined $T_0=T_{\pm}\big|_{\pazocal N}$. Taking this into account we state that if the test particle has approached the hypersurface from the past then the action of crossing the hypersurface is mathematically equivalent to making a coordinate transformation on the geodesic vector from the past flat coordinates, where for the purposes of our calculations we consider a trivial constant and parallel null congruence, to the continuous coordinate system. This transformed congruence then forms the initial conditions for the congruence to the future of the hypersurface.  

\be
\label{eq:contin}
	T_{+}^\alpha\big|_{\pazocal N}=\left.\left(\frac{\partial x_+^\alpha}{\partial x_-^\beta}T_-^\beta\right)\right|_{\pazocal N}
\ee

Here $T_{-}^\alpha$ is the geodesic vector of the test particle in the past domain in past flat coordinates and $T_{+}^\alpha$ is the corresponding vector after the particle crosses the shell in future coordinates as shown in both figures \ref{cts-lines} and \ref{discts-lines}. Here we should recall that all the information regarding the stress-energy tensor on the shell, which also means the effect that the shell will have on the congruence, is fully encoded in the definition of the soldering conditions and thus in the Jacobian of the soldering transformation.

Note that the geodesics are straight lines in the future and the past in the corresponding coordinate systems and with affine parameters $\lambda_{\pm}$ they are given by
\be
	x^\alpha_{\pm} = x_0^\alpha +  \lambda_{\pm} T_{\pm}^\alpha\big|_{\pazocal N}.
\ee
There is a one parameter freedom in the choice of affine parameters
\be
	\lambda_{\pm}\rightarrow\alpha^{-1}_{\pm} \lambda_{\pm},
	\qquad T_{\pm}^\alpha\big|_{\pazocal N} \rightarrow \alpha_{\pm} T_{\pm}^\alpha\big|_{\pazocal N} ,
\ee 
and the continuity equation \eqref{eq:contin} establishes a one-to-one relation between $\alpha_{-}$ and $\alpha_{+}$, thus fixing the affine parameter in the future we also fix the affine parameter in the past.

%================================================================================

\section{Null Congruences crossing horizon shells}
\label{sec:application}
Applying the proposed algorithm of the previous section we consider the
congruence $T_- = \alpha\partial_{u_-}$ globally to the past of ${\pazocal N}$ ($\alpha$ will be fixed after fixing the affine parameter to the future, as discussed in the previous section) and perform on ${\pazocal M^+}$ a coordinate transformation parametrised by $F(x^a)$ where $a=v,x,y$,
\be
	u_- = \frac{u}{F_v}, \qquad
	v_-= F + \frac{u}{2F_v}(F_x^2 + F_y^2),\qquad
	x_- = x + \frac{uF_x}{F_v},\qquad
	y_- = y + \frac{uF_y}{F_v}.
\ee
We will refer to this transformation as a Newman-Unti soldering being the extension to a soldering of the Newman-Unti transformation $v_- = F$.        
\begin{figure}[h!]
  \center{\includegraphics[width=7cm]{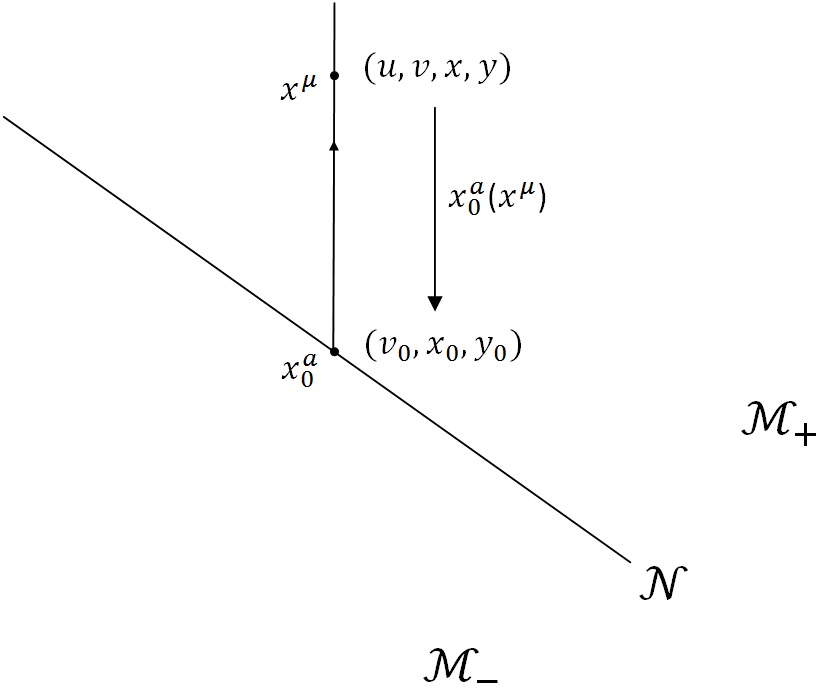}}
  \caption{Every point to the future of ${\pazocal N}$, apart from caustic points, has a unique mapping onto ${\pazocal N}$ obtained by following the geodesic of the congruence in ${\pazocal M^+}$ that passes through that point back to ${\pazocal N}$.}
  \label{maptoN}
\end{figure}
This one sided soldering transformation creates a shell at the location of  ${\pazocal N}$ and the properties of the shell are encoded in the function $F(x^a)$ as described in detail in \cite{Blau:2015nee}. To the future of ${\pazocal N}$ we have coordinates $x_+^\alpha = x^\alpha$ and we identify the future and past coordinates on ${\pazocal N}$. After the transformation the metric to the past of the shell is
\be
\label{eq:past_metric}
ds^2_-=-2\diff u \diff v +dx^2+dy^2+u\left(\frac{2}{F_v}F_{ab}dx^a dx^b\right)+\frac{u^2}{F_v^2}\left(F_{xa}F_{xb}+F_{ya}F_{yb}\right)dx^a dx^b,
\ee
while to the future it remains
\be
\label{eq:future_metric}
\diff s_+^2=-2\diff u \diff v+\diff x^2+\diff y^2.
\ee

We are interested in the value of the congruence $T_0$ on  ${\pazocal N}$ in continuous coordinates,
\be
T_0^\alpha=\frac{\partial x^\alpha}{\partial x_-^\beta}T_-^\beta\big|_{\pazocal N}
\ee
Inverting the Jacobian matrix of the coordinate transformation evaluated on ${\pazocal N}$ we find, for our choice of $T_-$, that 
\be
T_0(x_0^a) = \alpha\left.\left(F_v\partial_u + \frac{1}{2F_v}(F_x^2 + F_y^2)\partial_v - F_x\partial_x 
-F_y\partial_y\right)\right|_{\pazocal N}
\ee

The null congruences to the future of ${\pazocal N}$ are labeled by the point $(x_0^a)$ at which they cross ${\pazocal N}$ and the affine parameter $u$. Taking $T_0$ as the initial condition for the congruence on  ${\pazocal N}$ at $u=u_0=0$ we find that the null congruence to the future is described by the lines
\be
\label{eq:lines}
x^\alpha = x_0^\alpha +  u T^\alpha_0(x_0^a)
\ee
and from the $u$ component of this equation we find that
\be
	\alpha = 1/F_v.
\ee

The remaining components of \eqref{eq:lines} can in principal be inverted (in practice there will be unavoidable problems of caustics meaning that the inversion from some future points will not be well defined - we will ignore these subtleties), to obtain a projection along geodesic lines from ${\pazocal M^+}$ to ${\pazocal N}$ of the form $x_0^a = x_0^a(x^\alpha)$ as illustrated in figure \ref{maptoN}. The congruence to the future is then simply $T^\alpha(x^\mu) = T_0^\alpha(x_0^a(x^\mu))$. In the following, by a slight abuse of notation, we will use $F$ to denote the extension of the soldering transformation $F$ to the future such that $F(x^\alpha) = F(x_0^a(x^\alpha))$. A simple and useful consequence of this construction, that one can show with the help of the $a$ components of \eqref{eq:lines}, is
\be
\frac{\partial F}{\partial x^a} = \frac{\partial F}{\partial x_0^a},
\ee
With a little further work one can show that the congruence to the future of the shell is given by
\be
	\label{eq:orth_hyp}
	T^\mu = -\frac{1}{F_v}\ \eta^{\mu\nu}\partial_\nu F(x_0^a(x^\alpha)),
\ee
and is thus hypersurface orthogonal as anticipated at the end of section \ref{sec:memory}.

\section{How the shell modifies the congruence.}
\label{sec:behaviour}
We now turn to the projection of the B-tensor and its behaviour upon crossing the shell as described in section \ref{sec:memory}. A natural choice for completing the tetrad along the congruence is
\be
\label{eq:proj_vectors}
e_A = -\frac{F_A}{F_v}\partial_v + \partial_A
\ee
together with $n=\partial_v$ and the tangent vector $T$. We will also need the completeness relation
\be
e^\alpha_A e^\beta_B\delta^{AB} = \eta^{\alpha\beta} + n^\alpha T^\beta + n^\beta T^\alpha.
\ee
As already discussed a null shell produces a delta function singularity in the Riemann tensor and the physical content of the shell is encoded in the jump in the orthogonal derivatives of the metric tensor
\be
\gamma_{ab} = T^\alpha[\partial_\alpha g_{ab}] = -2\frac{F_{ab}}{F_v}|_\pazocal{N}\qquad \gamma_{u\alpha}=0.
\ee
The shell in general contains matter with stress-energy tensor
\be
S_{\alpha\beta} = \mu n_\alpha n_\beta + p g_{\alpha\beta} + 2j_{(\alpha}n_{\beta)}
\ee
with $j_\alpha = (0,j_a)$. The four independent components of the stress energy tensor are the energy density $\mu$, and the surface current $j_a$, the $v$ component of which is minus the pressure $p$
\be
\mu = -\frac{1}{16\pi}\gamma_{\alpha\beta}\eta^{\alpha\beta} \qquad j_a =\frac1{16\pi}\gamma_{a\beta} n^\beta\qquad p = -j_v =  -\frac{1}{16\pi}\gamma_{\alpha\beta}n^\alpha n^\beta.
\ee
These account for four out of the six independent components of $\gamma_{\alpha\beta}$, the remaining two coming from the spatial $(x,y)$ part of $\hat{\gamma}_{\alpha\beta}$
\be
\hat{\gamma}_{\alpha\beta} = \gamma_{\alpha\beta} - \frac12\gamma_{\delta\kappa}\eta^{\delta\kappa}\eta_{\alpha\beta}
\ee
which contribute to the Weyl tensor and encode the two polarisations of an impulsive gravitational wave on the shell. We will see in detail how this works below. 

To study the behaviour of a null congruence crossing the null shell we need to calculate $R_{AB}$ and it is straightforward to show that
\be
\bar{R}_{AB} = -\frac12\gamma_{\alpha\beta}e_A^\alpha e_B^\beta = -\frac12\gamma_{AB}.
\ee

Given the Einstein equation
\be
R_{\mu\nu} - \frac12 g_{\mu\nu}R = 8\pi S_{\mu\nu}\delta(u)
\ee
we can relate the trace of $\bar{R}_{AB}$ to the surface quantities
\be
\bar{\mathcal{R}} = 8\pi S_{\mu\nu}T^\mu T^\nu = 8\pi\mu - 16\pi j_aT^a,
\ee
while the projection of the Weyl tensor on the congruence is
\be
\bar{C}_{AB} = -\frac12\gamma_{AB} + \frac14\gamma^C_C\delta_{AB} = -\frac{1}{2}\hat{\gamma}_{\alpha\beta}e_A^\alpha e_B^\beta + 16\pi j_a T^a\delta_{AB}.
\ee

\subsection{Newman-Unti soldering transformations}

Taking the explicit form for $T^\mu$ from the previous section we find for a general Newman-Unti type transformation that
\be
\label{eq:general_B}
\pazocal{B}_{AB} = e_A^\alpha e_B^\beta \pazocal{B}_{\alpha\beta} = -\frac{F_{AB}}{F_v} - F_AF_B\frac{F_{vv}}{F_v^3} + \frac{(F_A F_{Bv} + F_B F_{Av})}{F_v^2}.
\ee
Evaluating $\pazocal{B}_{AB}$ on the shell gives us directly its discontinuity given that we have taken a congruence with $\pazocal{B}_{AB}=0$ before the shell. In this expression we must take care to recall that although $\partial_{a0}F = \partial_aF$ second derivatives must include the Jacobian of the mapping $x_0^a(x^\alpha)$. We see that $\pazocal{B}_{AB}$ is symmetric and thus the congruence has zero rotation consistent with the hypersurface orthogonality demonstrated in the previous section and also the more general arguments of section \ref{sec:memory}.

Evaluating explicitly $\bar{\mathcal R}$ and $\bar{C}_{AB}$ and comparing to \eqref{eq:general_B} we find that the change in expansion upon crossing the shell
\be
\label{eq:dtheta}
\theta|_\pazocal{N} = -\bar{\mathcal R} = - 8\pi\mu + 16\pi j_aT^a
\ee
is determined by a combination of the shell energy density and surface currents while the change in the shear
\be
\label{eq:dC}
{\sigma_{AB}}|_\pazocal{N} = -\bar{C}_{AB} = \frac{1}{2}\hat{\gamma}_{\alpha\beta}e_A^\alpha e_B^\beta - 16\pi j_a T^a\delta_{AB}
\ee
is determined by the gravitational wave component and surface current of the shell.

\subsection{BMS soldering}

To explicitly evaluate $\pazocal{B}_{AB}$ \eqref{eq:general_B} also to the future of the shell we need to invert equations \eqref{eq:lines} as discussed in the previous section. We will simplify the following calculations by just considering the special case of BMS supertranslation solderings and thus we take 
\be
\label{eq:bms_soldering}
F(v,x,y) = v + f(x,y).
\ee

Then
\be
\label{eq:bms_geodesic}
T_{\alpha} = -\partial_\alpha F = (-\frac12(f_x^2 + f_y^2),-1, -f_x,-f_y)
\ee
and
\be
\pazocal{B}_{AB} =  -\partial_B f_A = - \frac{\partial x_0^C}{\partial x^B}\frac{\partial f_A}{\partial x_0^C}.
\ee
In this case we need only the Jacobian of the transformation on spatial coordinates that we obtain by taking derivatives of the $x,y$ components of \eqref{eq:lines} with respect to $x_A = (x,y)$ to obtain
\be
\delta_A^B = \frac{\partial x_0^C}{\partial x^A}(\delta_C^B - u f_{BC})
\ee
and inverting we find the Jacobian of the transformation
\be
\left(\frac{\partial x_0^B}{\partial x^A}\right) = \frac{1}{1 - u\trf + u^2\detf}
\begin{pmatrix}
     1-uf_{yy}&uf_{xy} \\
     uf_{xy} &1-uf_{xx} 
\end{pmatrix},
\ee
where  $\trf=f_{xx}+f_{yy}$ and $\detf = f_{xx}f_{yy} - f_{xy}^2$.
Thus
\be
	\pazocal{B} =\frac{-1}{1 - u\trf + u^2\detf}
	\begin{pmatrix}
		f_{xx} - u\detf &f_{xy} \\
		f_{xy} & f_{yy} - u\detf
	\end{pmatrix}
\ee
corresponding to the expansion 
\be
\label{eq:expansion} 
	\theta =\frac{-\trf + 2 u\,\detf}{1 - u \trf + u^2 \detf}
\ee
and shear
\be
\label{eq:shear}
	\sigma =\frac{-1}{2(1 - u \trf + u^2 \detf)} 
        \begin{pmatrix}
    	f_{xx}-f_{yy}&f_{xy} \\
    	f_{xy}& -f_{xx}+f_{yy}
	\end{pmatrix}.
\ee
Evaluating
\be
\bar{\mathcal R} = f_{xx} + f_{yy} = 8\pi\mu\qquad
\bar{C} = \frac12\begin{pmatrix}
    	f_{xx}- f_{yy}&2f_{xy} \\
    	2f_{xy} & -f_{xx}+f_{yy}
	\end{pmatrix} = -\frac12\hat{\gamma}
\ee
it is easy to check that our solutions for expansion and shear on and to the future of $\pazocal{N}$ satisfy the evolution equations
\be
	\frac{\dd \theta}{\dd u} = - \frac12 \theta^2 - 2(\sigma_+^2 + \sigma_\times^2) - 8\pi\mu\delta(u)
	\qquad
	\text{and}
	\qquad
	\frac{\dd \sigma}{\dd u} = -\theta  \sigma +\frac12\hat{\gamma}\delta(u).
\ee

We see in particular that for the BMS transformations the $\pazocal{B}$-memory effect corresponds to a jump in the expansion upon crossing the shell that is proportional to the energy density of the shell together with a change in the shear that is proportional to the gravitational wave component of the shell. 

\section{Discussion}
\label{sec:discussion}

We have presented a new approach for studying congruences that cross a singular hypersurface. Our method is based on the physically justified assumption that the geodesic vector of a test particle is continuous across the hypersurface when using continuous coordinates. To obtain the geodesic flow to the future of the hypersurface one simply needs to do a coordinate transformation on the past coordinates to go to a continuous coordinate system. The resulting transformation on the geodesic congruence in $\pazocal M^-$ gives initial conditions on $\pazocal N$ to develop the geodesic vector field on $\pazocal M^+$ to the future.

We then proved that a parallel congruence upon crossing the shell gives rise to a hypersurface orthogonal congruence to the future of the shell, and in particular that the shell gives rise to a discontinuity in the $\pazocal{B}$-tensor of the congruence. In general the jump in the expansion is determined by the energy density and currents on the shell while the jump in the shear is determined by the gravitational wave component together with the surface currents. Although we derived these results using a particular congruence, it should be clear from \eqref{eq:dtheta} and \eqref{eq:dC} that the results are independent of the choice of congruence in the case of BMS supertranslations for which the surface currents are zero. We also provide a general argument that a hypersurface orthogonal congruence before the shell will give rise to a hypersurface orthogonal congruence to the future.

The change in the $\pazocal{B}$-tensor after the passage of an outgoing gravitational wave leads to a covariant description of the gravitational memory effect - the $\pazocal{B}$-memory effect.
Although our construction and approach to gravitational memory appears to be quite distinct from that reviewed in \cite{Ashtekar:2018lor} there are many intriguing similarities. They introduce a trace free ``shear like'' tensor
$\sigma_{ab} = \nabla_a\nabla_b f$ where $f$ is the shift in a BMS supertranslation
on $\pazocal{I}$ and the Lie derivative along $\pazocal{I}$ of $\sigma_{ab}$ is the news tensor $N_{ab}$  \cite{Bondi:1962px}. The picture that emerges suggests that the outgoing null shell induces a BMS supertranslation on $\pazocal{I}$ in the same way that a soft graviton is supposed to \cite{Strominger:2017zoo}. 

It would be very interesting to study the quantum version of this effect and the calculation of the eikonal wavefunction may be a first step in such an approach. In the eikonal picture the local wavefronts of a wavefunction follow the geodesics of the spacetime. The presence of an outgoing gravitational wave produces a radical reorganisation of the congruence such that in general a flat wavefront can be distorted in a myriad of different ways. One may imagine that at a deeper level this distortion corresponds to a radical change in the quantum field theory vacuum that is constructed from plane wave states. It would be interesting in particular to investigate how the propagation across the shell of a good basis of wave-functions may not give rise to a reasonable basis to the future of the shell given that BMS transformations map between inequivalent quantum field theory vacuum states \cite{Ashtekar:2018lor}. 

%===============================================================================

\acknowledgments
MO thanks M. Blau for discussions and for various comments and suggestions. MO would also like to thank the A.~I.~Alikhanian National Science Laboratory and the Ambartsumian Byurakan Astrophysical Observatory for hospitality during the initial stages of this work. The work of HD is supported in part by the Armenian State Committee of Science, Grant No. 18RF-002, by VolkswagenStiftung, in part by research grant form the Armenian National Science and Education Fund (ANSEF) based in New York, USA and the Foundation for Armenian Science and Technology (FAST). We acknowledge the ICTP Affiliated Center program AF-04.

\end{document}